\documentclass[twocolumn,superscriptaddress,amsmath,amssymb, aps,prx]{revtex4-2}

\usepackage{graphicx}
\usepackage{mathtools}
\usepackage{multirow}
\usepackage{array}
\usepackage{hyperref}

\usepackage{xcolor}
\definecolor{myblue}{RGB}{29, 113, 184}
\definecolor{myyellow}{RGB}{254, 145, 0}

\hypersetup{
	colorlinks=true,
	linkcolor=myyellow,
	filecolor=magenta,      
	urlcolor=cyan,citecolor=myblue}
\usepackage[nohyperlinks, printonlyused]{acronym}
\usepackage[capitalise]{cleveref}

\newcommand{\hvalues}{{$h$-values}}
\newcommand{\hvalue}{{$h$-value}}

\newcommand{\gmatrix}{{$\mathcal{G}$}}

\begin{document}
	
\title{Optimal  foraging strategies can be learned }

\author{Gorka Mu\~noz-Gil}
\thanks{These authors contributed equally to the work}
\author{Andrea L\'opez-Incera}
\thanks{These authors contributed equally to the work}
\author{Lukas J.~Fiderer}
\author{Hans J.~Briegel}
\affiliation{Institute for Theoretical Physics, University of Innsbruck, Technikerstr. 21a, A-6020 Innsbruck, Austria}

\begin{abstract}
The foraging behavior of animals is a paradigm of target search in nature. Understanding which foraging strategies are optimal and how animals learn them are central challenges in modeling animal foraging. While the question of optimality has wide-ranging implications across fields such as economy, physics, and ecology, the question of learnability is a topic of ongoing debate in evolutionary biology. Recognizing the interconnected nature of these challenges, this work addresses them simultaneously by exploring optimal foraging strategies through a reinforcement learning framework. To this end, we model foragers as learning agents. We first prove theoretically that maximizing rewards in our reinforcement learning model is equivalent to optimizing foraging efficiency. We then show with numerical experiments that, in the paradigmatic model of non-destructive search, our agents learn foraging strategies which outperform the efficiency of some of the best known strategies such as L\'evy walks.  These findings highlight the potential of reinforcement learning as a versatile framework not only for optimizing search strategies but also to model the learning process, thus shedding light on the role of learning in natural optimization processes.
\end{abstract}

\maketitle

Strategies to search for randomly distributed targets are of paramount importance in many fields. For instance, they are widely used in ecology to model the foraging activities of predators~\cite{sims2008scaling}, hunters~\cite{raichlen2014evidence} and gatherers~\cite{brown2007levy}, as well as the movement of pedestrians in complex environments~\cite{murakami2019levy}; other interesting applications are the study of human information search in complex knowledge networks~\cite{lydon2021hunters} or the improvement of optimization algorithms~\cite{yang2009cuckoo}.

\begin{figure}[t]
\includegraphics[width=\columnwidth]{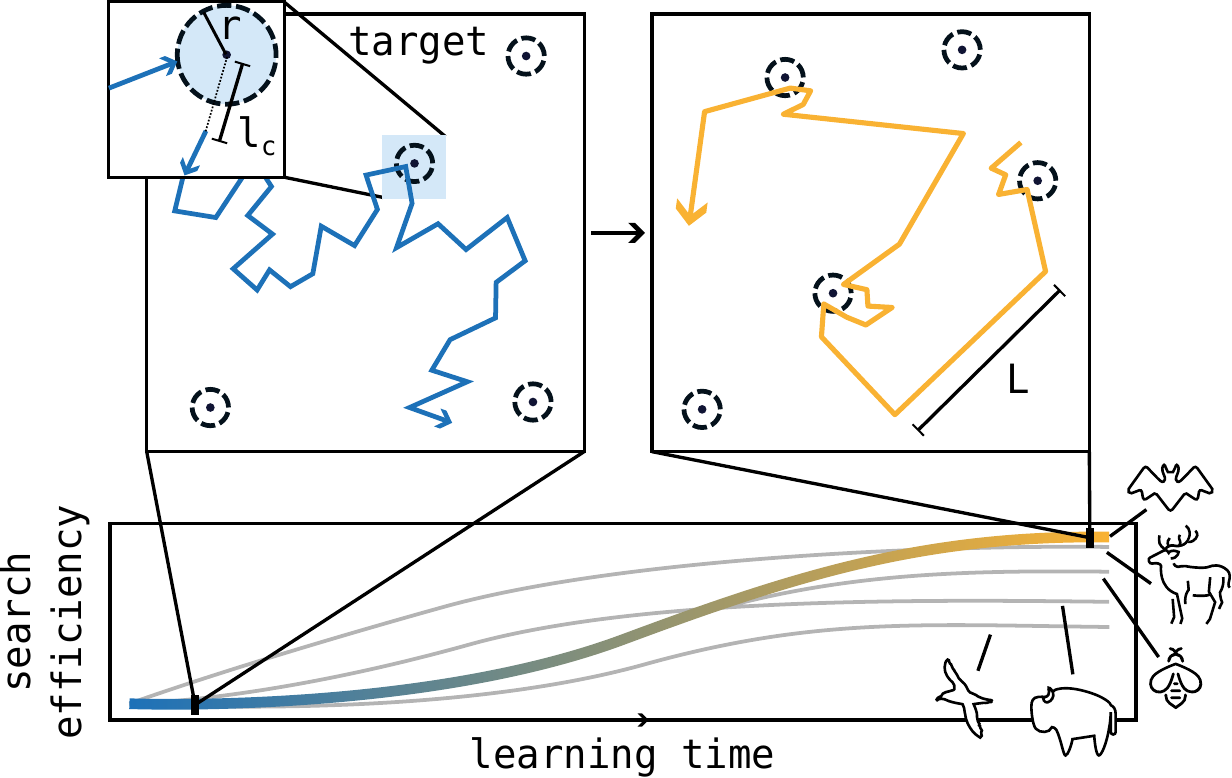}
\caption{Schematic illustration of random-walk based foraging strategies. According to the hypothesis of learning-based movement biological entities can adapt and learn in order to optimize their search efficiency (i.e. the number of targets collected per time)~\cite{lewis2021learning}. Different animals (as e.g. albatrosses~\cite{viswanathan1996levy, de2017early}, bison~\cite{sigaud2017collective}, bumblebees~\cite{lihoreau2012radar, leadbeater2009bumble}, deer~\cite{focardi2009adaptive} or bats~\cite{vilk2022phase}) may learn different strategies, based on their cognitive capacities, surrounding environment or biological pressure~\cite{wosniack2017evolutionary}.}
\label{fig:fig0}
\end{figure}

A widely used and investigated idealization of such search problems is the model of non-destructive foraging (cf.~\cref{fig:fig0}). The model consists of a two-dimensional environment with randomly, uniformly, and sparsely distributed immobile targets which can be detected by a walker within a \emph{detection radius} $r$. The walker moves through the environment with constant speed in steps (straight segments) of random orientation and varying \emph{step length} $L$. As soon as the walker detects a target, its current step is aborted, whereupon it immediately takes its next step from a position displaced in a random direction by a \emph{cutoff length} $l_\textrm{c}> r$ from the detected target.
	
The goal of the walker is to maximize the search efficiency 
\begin{align}
	\eta = \lim_{T\rightarrow \infty}\frac{n_T}{T} \label{eq:efficiency}
\end{align}
where $T$ is physical time and $n_T$ is the average number (with respect to target positions) of detected targets in the time interval $[0,T]$. In several fields, $\eta^{-1}$ is known as the mean first-passage time (MFPT) since it equals the average time a walker needs to detect the first target \cite{levernier2020inverse}. Indeed, MFPT-based approaches are widely used to study and compare different search strategies~\cite{palyulin2014levy, rupprecht2016optimal, chechkin2018random, palyulin2017comparison}. However, the hardness of computing the MFPT in realistic scenarios poses an obstacle to the theoretical understanding of optimal search strategies.

Non-destructive foraging corresponds to the biologically relevant case of foraging replenishable targets where larger cutoff lengths $l_\textrm{c}$ correspond to longer replenishment times of targets \cite{raposo2003dynamical}. In particular, the case of \textit{destructive} foraging, where targets vanish after detection, can formally be recovered by taking the limit $l_\textrm{c}\rightarrow \infty$. 

In the foraging model under consideration, simplifying assumptions are made regarding the walker's selection of step length. These assumptions state that the walker's choice of step length is independent of its position, any previous detection events, and also any prior choices of step lengths. This results into a widely investigated class of random walks, which considers that the walker samples independently from a probability distribution $\Pr(L)$ at each step to determine its current step length. This is the class of random walks we consider in this work.

Motivated by observational data \cite{cole1995fractal, schuster1996chemosensory, viswanathan1996levy} and supported by heuristic arguments \cite{levandowsky1988swimming, viswanathan1996levy, viswanathan1999optimizing}, a family of step-length distributions known as L\'evy distributions~\cite{shlesinger1995levy, zaburdaev2015levy} $\Pr(L)\propto s^\mu L^{-(\mu+1)}$,
where $s$ is a length-scale parameter and $\mu\in[0,2]$, received major attention, including the discovery of the optimal value of $\mu$ in certain limiting cases \cite{viswanathan1999optimizing, levernier2020inverse}, such as $l_\textrm{c}\rightarrow \infty$ and $l_\textrm{c}\rightarrow r$. While a number of numerical studies have shown that typically the optimal value of $\mu$ depends sensitively on the environmental parameters \cite{viswanathan1999optimizing, raposo2003dynamical, santos2004optimal, volpe2017topography, levernier2020inverse}, it has recently been shown that other families of step-length distributions $\Pr(L)$ such as bi- and tri-exponential distributions outperform L\'evy distributions~\cite{tejedor2012optimizing, palyulin2014levy, benhamou2015ultimate, ferreira2021landscape}. In general, it is still an open problem what strategy is optimal for non-destructive foraging, both in terms of its step-length distribution family and its precise parameters.
	
Apart from the exact mathematical description of the search strategies followed by biological entities, another debate concerns the question of how animals come to possess knowledge of particular search strategies~\cite{lewis2021learning}, which recently sparked interest from a neurological perspective~\cite{jones2022scale, sims2019optimal}. More generally, the origin of search strategies is usually classified as either \emph{emergent} or \emph{evolutionary}~\cite{reynolds2015liberating}. The former considers that an individual walker can learn \emph{different} strategies via complex interactions with different environments, while the latter proposes that a \emph{single} strategy has evolved via natural selection in order to optimize search in different environments~\cite{wosniack2017evolutionary}. In order to understand emerging and adaptive processes one has to go beyond the study of optimal search strategies based on heuristic ansätze; it calls for a framework that naturally incorporates learning.

This paper presents such a framework, rooted in the principles of reinforcement learning (RL). What sets our model apart from normative approaches, which aim to directly optimize an objective function such as search efficiency, is its mechanistic nature~\cite{levenstein2023role}: in the proposed RL framework, optimization occurs as a result of an interactive learning process which does not assume direct access to the objective function. This allows for valuable insights into the optimization of search strategies and the comprehension of their emergence through the process of learning. 
In summary, by abstracting away extraneous biological details, we arrive to a concise framework specifically designed to model how biological agents learn to forage.
\begin{figure}[t]
\includegraphics[width=\columnwidth]{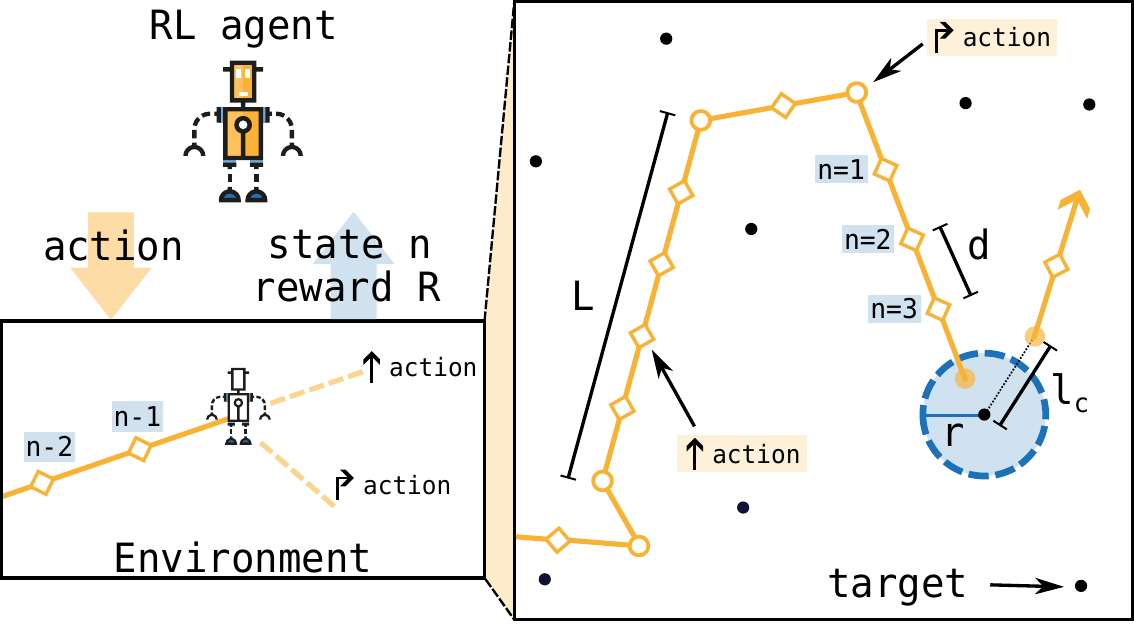}
\caption{The problem of non-destructive foraging is formulated within the framework of RL. An agent moves through an environment with randomly distributed targets and, at each step, chooses between two possible actions: continue in the same direction ($\uparrow$) or turn ($\Rsh$) in a random direction. The state perceived by the agent is a counter $n$, which is the number of small steps of length $d$ which compose the current step of length $L$. Whenever the agent detects a target, it receives a reward $R$ and resumes its walk at a distance $l_\mathrm{c}$ from the detected target with the counter reset.}
\label{fig:scheme}
\end{figure}

\emph{Non-destructive foraging in the RL framework.---}\label{sec:agent and env}
We base our framework on RL, a standard paradigm of machine learning~\cite{sutton2018reinforcement}. To this end we will first describe RL in general before we proceed with embedding non-destructive foraging within the framework of RL.

RL considers an agent interacting with an environment (see \cref{fig:scheme}). During each interaction cycle, called an \emph{RL step}, the agent perceives \emph{state} $s\in\mathcal{S}$ from the environment, and replies with an \emph{action} $a\in\mathcal{A}$. We adopt the common assumption that the agent's decision making is Markovian, which means that the probability $\pi(a|s)$ that the agent chooses action $a$ only depends on the current state $s$. $\pi(a|s)$ is called the \emph{policy} of the RL agent.  Further, we consider the environment to be partially observable, which means that the agent cannot perceive the full state $e\in\mathcal{E}$ of the environment, that is $s\neq e$.
 
The agent also receives a \emph{reward} signal $R(e, a)$ depending on the current action $a$ and the state of the environment $e$. The goal of the agent is then to optimize its policy such that the expected average reward
\begin{align}
	R_\pi(e) = \lim\limits_{T\rightarrow \infty}\frac{\mathbb{E}\left[\sum_{t=0}^{T}  R(e_t, a_t)\bigg\rvert \pi, e_0=e\right]}{T} \label{eq:reward}
\end{align}
is maximized, where $t\in\mathbb{N}_0$ labels the RL steps and $\mathbb{E}$ denotes expectation with respect to all infinite trajectories of agent-environment interactions given that the agent employs policy $\pi$ and that the environment is initially in state $e\in\mathcal{E}$. In practice, the optimization of \cref{eq:reward} with respect to the policy $\pi$ is governed by an RL algorithm which will be discussed further below.

In the following, we embed non-destructive foraging within the framework of RL. The idea is to model the walker as an RL agent. A naive approach would be to let the agent choose its step length $L$ in each RL step without ever perceiving a state from the environment. However, this would correspond to an infinite action space $\mathcal{A}$ and an empty state space $\mathcal{S}$,  which often represents an obstacle to efficient learning.

Therefore, we proceed by reformulating the non-destructive foraging model. As we will show below, this is only a reformulation and hence will give rise to exactly the same class of walks which are fully characterized by sampling independently at each step from a step-length distribution $\Pr(L)$. The reformulation consists in discretizing the step length $L$ in units of \textit{small steps} $d$ such that at each RL step the agent only has the following choice: either continue in the direction of its previous step or \textit{turn} in a random direction.
This formulation indeed resembles that of Ref.~\cite{tejedor2012optimizing}, although learning was not considered in that work.
 
To this end, given the environmental parameters $r$, $l_\textrm{c}$, and $d$, which characterize the search problem, the state of the environment is defined by the positions of targets, together with the current position, walking direction, and a \emph{step counter} $n\in\mathbb{N}$ of the agent. The counter is reset to $n=1$ whenever the agent turns or detects a target and it increases by one whenever the agent continues without detecting a target. In each RL step, the counter is perceived by the agent, i.e., $s=n$. The two possible actions are \emph{continue} and \emph{turn}, symbolically denoted by $\uparrow$ and $\Rsh$, which correspond to walking for a distance $d$ either in the current or in a random direction, respectively. If the agent detects a target, its position is resampled at a distance $l_\mathrm{c}$ from the detected target, according to the non-destructive foraging model. Then, the agent walks a small step of length $d$ in a random direction, and it perceives the counter state $n=1$ and a reward $R=1$. Otherwise, when no target is detected, $R=0$ and the agent continues its walk unimpeded.

The probability that the agent makes a step of length $L$ is thus given by
\begin{align}
\Pr(L = Nd) = \pi(\Rsh|N) \prod_{n = 1}^{N-1} \pi(\uparrow|n),
\label{eq:prob}
\end{align}
where we use the convention that the empty product, in case of $N=1$, equals one.  Since there are only two actions, we have $\pi(\Rsh|n) = 1-\pi(\uparrow|n)$.
  
We proceed by showing that our RL agent maximizes its search efficiency $\eta$ by maximizing the expected average reward in \cref{eq:reward}. First we note that the average number of targets detected during the first $T$ RL steps, given that policy $\pi$ is employed and that the environment is initially in state $e$, can be written as
 \begin{align}
 	n_{T} = \mathbb{E}\left[\sum_{t=0}^{T}R(e_{t},a_{t})\bigg\rvert \pi, e_0=e\right], \label{eq:number}
 \end{align}
since $R=1$ when a target is detected and $R=0$ otherwise. By inserting \cref{eq:number} into the definition of the detection efficiency, \cref{eq:efficiency}, we immediately see that $\eta = R_\pi (e)$ and therefore, $\eta$ is maximal when $ R_\pi(e)$ is. This proves an important feature of our RL formulation of foraging: reinforcement learning converges to optimal search strategies.
 
Since our RL formulation moves the focus from step-length distributions $\Pr(L)$ to policies $\pi$, we proceed to show which policy will reproduce a given step-length distribution.  Policy and step-length distribution are related via \cref{eq:prob}. Inserting $\pi(\uparrow|N) = 1-\pi(\Rsh|N)$ into \cref{eq:prob} yields a recurrent equation for $\pi(\Rsh|n)$, the solution of which we find to be
	\begin{align}
		\pi(\Rsh|N)= \frac{\Pr(L = Nd)}{1-\sum_{n=1}^{N-1}\Pr(L = nd)}% = \frac{1-\sum_{n=1}^{N}\Pr(L = nd)}{1-\sum_{n=1}^{N-1}\Pr(L = nd)}
		 \label{eq:solution}
	\end{align}
for any $N$ such that the denominator is non-vanishing, i.e., $\sum_{n=1}^{N-1}\Pr(L = nd) \neq 1$, or equivalently, $\sum_{n=N}^{\infty}\Pr(L = nd) \neq 0$. If $\sum_{n=N}^{\infty}\Pr(L = nd) = 0$, no step can continue beyond length $(N-1)d$, which implies $\pi(\Rsh|n)=1$ for an $n<N$. In this case, the agent would always turn before perceiving states with $n\geq N$ and the policy for $n\geq N$ does not affect the walk.

Some important remarks are in order: (i) while the agent has information on the current step-length via the counter state, it is unaware of any previous step lengths, due to the counter reset; (ii) given the former, the motion of our agent classifies as a semi-Markov process since the points where direction changes occur form a Markov chain, as it is the case for most L\'evy walks and related diffusion processes~\cite{zaburdaev2015levy}; (iii) last, as shown by \cref{eq:prob} and \cref{eq:solution}, the family of walks which can be generated by the RL agent is precisely that with $\Pr(L)$-distributed step lengths. In the form presented in this work, agents are not designed to and are not capable of performing walks that are non-Markovian in step length (walks that are non-Markovian in step length would be of the form $\Pr(L_{t} | L_{t-1},L_{t-2},...)$).

Finally, from \cref{eq:solution} it can also be seen that the agent must employ a probabilistic policy to reproduce arbitrary step-length distributions; deterministic policies with $\pi(s|a)\in\{0,1\}$ can only generate walks with constant step lengths.

\begin{figure}[t]
   \includegraphics[width=\columnwidth]{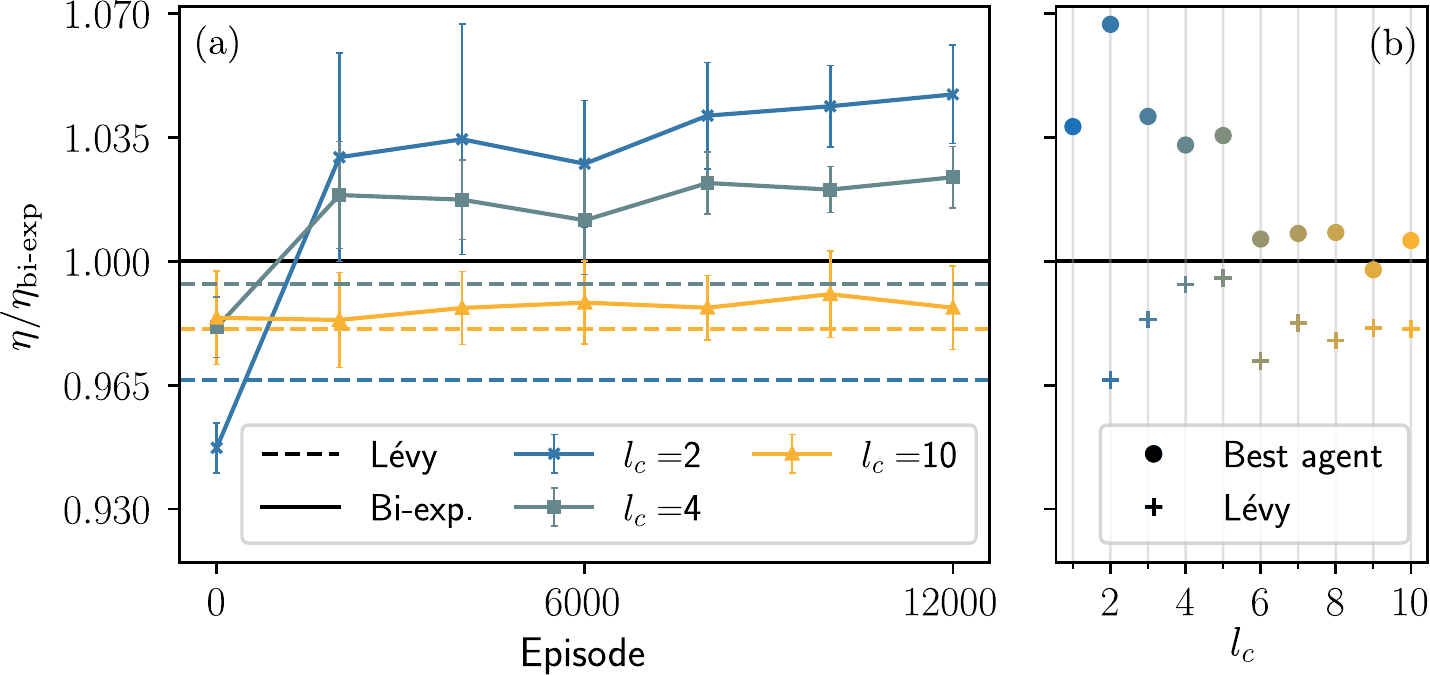}
   \caption{
   Learning curves and the advantage of learned policies over benchmarks. (a) The search efficiency (averaged over 10 agents, displayed with one standard deviation) is shown over the course of learning (measured in training episodes). Different colors correspond to environments with different cutoff length $l_\mathrm{c}$. Efficiencies are normalized by the respective best benchmark efficiency, which turns out to be bi-exponential distributions in all cases. Dashed lines show the efficiency of the best L\'evy walk for each case. (b) Comparison between the best agent's search efficiency at the end of learning and that of the best benchmarks, for each environment. The efficiency of the best L\'evy walk for $l_\mathrm{c}=1$ is $\eta_{\textrm{L\'evy}} / \eta_\textrm{bi-exp.} = 0.88$. 
   For each agent and benchmark model, the efficiency is averaged over $2\cdot10^8$ RL steps. In panel b), the standard error of the mean for the benchmark models and the best agents is depicted but too small to be visible.}
    \label{fig:efficiency}
\end{figure}
\emph{Numerical case study.---}
In the framework of RL, learning consists of updating the policy in order to maximize the obtained rewards (\cref{eq:reward}) in response to agent-environment interactions. From a plethora of RL algorithms, we choose projective simulation (PS)~\cite{briegel2012projective, Lopezincera20} (for details, see \cref{app:PS} and \cite{ourRepo}). We expect that similar results can be found with other RL algorithms such as policy gradient \cite{sutton2018reinforcement}.

As benchmarks for the trained RL agents, we use the widely investigated L\'evy walks \cite{shlesinger1995levy, viswanathan1996levy, viswanathan1999optimizing, levernier2020inverse} as well as the strongest ansatz considered in the literature so far: a random walk with bi-modal exponential step length distribution~\cite{palyulin2014levy, benhamou2015ultimate, ferreira2021landscape}(see \cref{app:benchmark_models} for details). Their search efficiencies are computed numerically by simulating an agent with an equivalent, fixed policy, obtained via \cref{eq:solution}, which for short will be referred to as L\'evy and bi-exponential policies, respectively.

% %%%%%%%%%%%%%%%%%%%%%%% Fake training %%%%%%%%%%%%%%%%%%%%%%%%%%%%%%%%%
It is important to note that an RL agent is not expected to learn any of the benchmark policies unless these are the only ones that maximize search efficiency (i.e. these strategies are indeed optimal). Nevertheless, we want to demonstrate, from a numerical perspective, that an RL agent based on PS has the capacity to learn any of these policies. To this end, we train agents to imitate the benchmark policies with various parameters and find a near-perfect convergence of their learned policies (for further details on imitation learning, see \cref{app:imitation}). These results suggests that, if any of the known foraging strategies is optimal, we can expect a trained agent converging to it. With this result in mind, we return to the original problem of learning optimal foraging strategies by interacting with an environment with randomly distributed targets, see \cref{fig:scheme}.

\begin{figure}[t!]
\includegraphics[width=0.8\columnwidth]{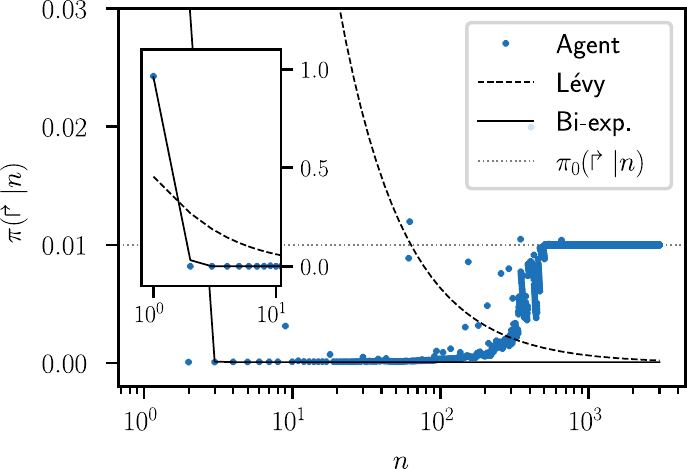}
   \caption{Policy of an RL agent trained in an environment with cutoff length $l_\mathrm{c}=0.6$. The policies corresponding to the best bi-exponential (solid line, $d_1=0.15, \omega_1=0.96, d_2=13047.89$) and the best L\'evy distributions (dashed line, $\beta=0.64$) are shown for comparison. The normalized search efficiencies are $\eta_{\textrm{RL}}/\eta_{\textrm{bi-exp}} = 1.02$ and $\eta_{\textrm{L\'evy}}/\eta_{\textrm{bi-exp}}=0.85$. The grey dotted line marks the initialization policy $\pi_0(\Rsh|n)=0.01$ $\forall n$.  
   }
    \label{fig:chosen_policy}
\end{figure}

%%%%%%%%%%%%%%%%%%%%%%% Pure numerics %%%%%%%%%%%%%%%%%%%%%%%%%%%%%%%%%
We model the infinite foraging environment by a two-dimensional squared box of size $W=100$ with periodic boundary conditions,  100 randomly distributed targets (thus corresponding to a target density $\rho = 0.01$) with a detection radius $r = 0.5$, where the unit length is defined as the small-step size of the agent $d=1$. In what follows, we consider that all lengths are measured in units of $d=1$ and we adopt a dimensionless notation. In this work, the agents are always initialized with a policy $\pi_0(\Rsh|n)=0.01~  \forall n$. Further details on this and the respective parameters of the learning process are presented in \cref{app:PS}.

\begin{figure*}[htb!]
   \includegraphics[width=\textwidth]{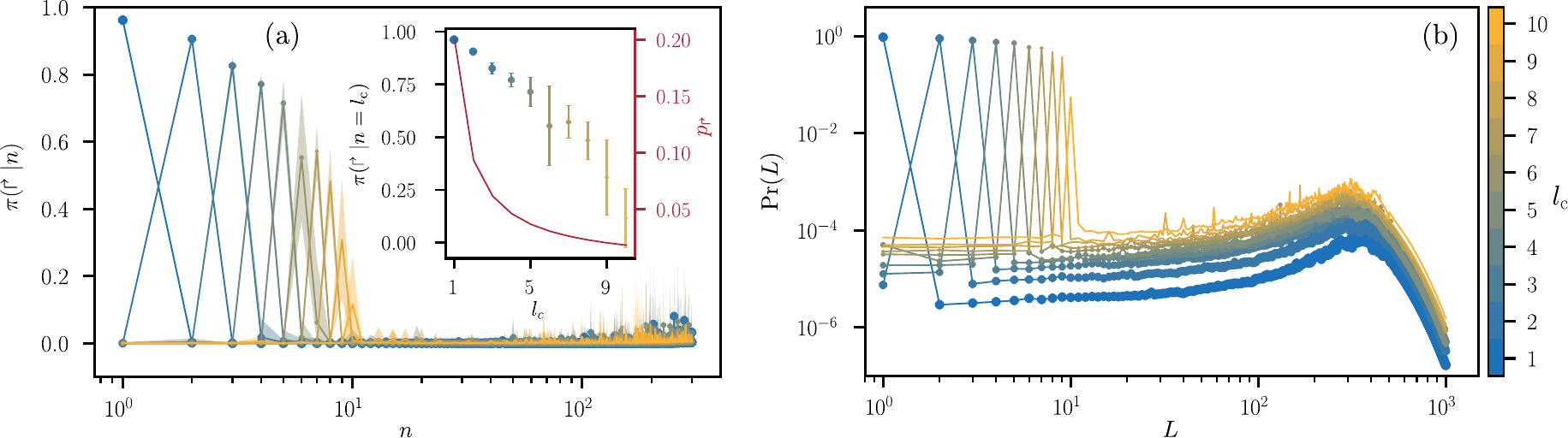}
   \caption{Analysis of the learned policies for different cutoff lengths.
   (a) Learned policies $\pi(\Rsh|n)$ as a function of the counter $n$. Each point is the average over 10 agents and the shaded area represents one standard deviation. 
   The inset shows, on the left axis, the turning probability at $n=l_\mathrm{c}$ averaged over 10 agents, error bars represent one standard deviation. On the right axis, the probability $p_\Rsh$ of hitting the target when turning at $n=l_\mathrm{c}$ is shown (see \cref{app:geometry}).
   (b) Step-length distributions corresponding to the policies presented in a). Each point is the median over 10 agents.
   }
    \label{fig:policies_and_dist}
\end{figure*}

We simulate RL agents in various foraging environments with different cutoff length $l_\mathrm{c}$. \cref{fig:efficiency} shows the learning curves of the agents, i.e., the evolution of search efficiency over the course of learning (see \cref{app:learning} for details).
The performance averaged over 10 agents reaches or overcomes the best benchmark in most cases, showing the robustness of the proposed RL method. Importantly, in all but one of the tested environments, the best of the 10 trained agents achieves higher efficiency than the respectively best benchmark model (see \cref{fig:efficiency}b), suggesting that there exist optimal walks with step-length distribution $\Pr(Nd)$ as given by \cref{eq:solution}, beyond the benchmark model families. 

In order to better understand the learned policies and the origin of such an advantage, we start by comparing the learned policy with the benchmarks at the example of $l_\mathrm{c} = 0.6$, see \cref{fig:chosen_policy}. For $n \lesssim 10^2$, training significantly changes the initial policy% ($\pi_0(\Rsh|n) = 0.01~\forall n$)
, and the learned policy shows greater similarity to the bi-exponential than to the L\'evy policy. For $n \gg 10^2$, the learned policy still equals its initial value of $0.01$. This is due to practical limitations of the training and can in particular be attributed to a limited training time, as detailed in \cref{app:policy_init}. However, note that the policy for large $n$ only affects the agent's behavior at sufficiently long steps that happen very rarely, and thus the resulting effect on the agent's walk is rather small.

\cref{fig:policies_and_dist}a shows the learned policies for environments with $l_\mathrm{c}\in [1,10]$, revealing an interesting property of the learned policies:
the probability of turning at $n=l_\mathrm{c}$ is significantly increased for all environments considered, and the smaller $l_\mathrm{c}$ the more pronounced this effect is (see inset of \cref{fig:policies_and_dist}a). This can be understood by an approximation: in the regime of low target density and small $l_\mathrm{c}$, after the detection of a target,  to a good approximation the immediate environment of the agent only contains the previously detected target. Then, apart from the possibility of returning to the target by walking straight for $l_\mathrm{c}$, there is a significant probability $p_\Rsh$ that the agent will first miss the target if it walks straight for $l_\mathrm{c}$ but then detects the target in the next step by turning at $n=l_\mathrm{c}$, see inset of \cref{fig:policies_and_dist}a. This explains the added benefit of turning at $n=l_\mathrm{c}$ (for details, see \cref{app:geometry}).

Using \cref{eq:prob}, we translate the learned policies from \cref{fig:policies_and_dist}a into the their corresponding step-length distributions (\cref{fig:policies_and_dist}b).
The learned distributions show a pronounced peak at $L=l_\mathrm{c}$, which can be interpreted as the agents learning to forage with a length scale related to the cutoff length. Then, for $10^1\lesssim L\lesssim 10^2$, we find a region where the probability remains stable, i.e., with a smaller probability the agents also make longer steps, associated with a second and larger length scale. For $L > 10^2$, $\Pr(L)$ increases before it decreases exponentially, which is due to the finite training times (see \cref{app:policy_init}). This prevents us from fully characterizing the second length scale (and any other that might appear for larger $L$), which is commonly related to the average spacing between targets~\cite{ferreira2021landscape}.

Indeed ``two-scale" foraging strategies, such as the bi-exponential benchmark, have been widely studied in the literature~\cite{benhamou2015ultimate, reynolds2015liberating, wosniack2017evolutionary} and their approximate emergence from our learning model is interesting.
It should be noted, however, that this emergence is only approximate: first, we found the learned policies to perform even slightly better than the bi-exponential benchmark in most cases. Second, neither the learned step-length distributions are fully characterized by two scales, nor is the problem of non-destructive foraging, since the detection radius $r$ represents a third length scale~\cite{ferreira2021landscape}. Therefore there are good reasons to hope for learning advantages over benchmark model families, as found in this work.

\emph{Discussion.---} 
We introduced a framework based on reinforcement learning (RL) for non-destructive foraging in environments with sparsely distributed immobile targets. The framework is based on analytic results which involve an exact correspondence between the policy $\pi(a|n)$ and the step-length distribution $\Pr(L)$ of the agent, and a proof that learning converges to optimal foraging strategies. This not only makes the proposed RL framework a viable alternative to traditional approaches, but also introduces remarkable benefits.

Firstly, learning is not restricted to a specific ansatz, as e.g. step-length distributions such as L\'evy walks or bi-exponential ansätze, which have been extensively studied in the literature. Even in cases where the latter are optimal, RL could be leveraged to find the suitable parameters of the respective model family. Most importantly, RL together with \cref{eq:prob} paves the way for the discovery of novel, more efficient step-length distributions.

Secondly, the RL framework sidesteps the problem of what knowledge about the environment (e.g., characteristic length scales) should be considered available to the agent; while it is often assumed that foragers do not possess any knowledge about the environment, it is also well-known that informed agents usually outperform their agnostic counterparts \cite{ferreira2021landscape, wosniack2017evolutionary}. With RL, it is not necessary to artificially endow the agent with knowledge about the environment. Instead, the agent implicitly \emph{learns} such knowledge by interacting with the environment. Interestingly, recent work has also shown that RL agents create implicit maps of their environments even when they are not directly programmed to do so \cite{wijmans2023emergence}. As shown in our numerical case study, RL agents are able to adapt to different length scales of the environment. Therefore, the assumption that foragers do not possess any knowledge about the environment is unrealistic for adaptive, learning foragers.   

The proposed framework could also be leveraged to understand how search strategies are learned by foragers in biological scenarios. This is related to the debate on the emergent versus evolutionary hypotheses (see Introduction). Both evolutionary priors and learning could be modelled within our framework via policy initialization and policy updating, respectively. For example, we showed numerically (\cref{fig:efficiency}) that an agent initialized with a preference for walking straight efficiently learns to perform better than L\'evy searches. 

Moreover, we expect RL to be beneficial especially when dealing with more complex environments, where exploiting the numerous properties of the environment can lead to better foraging strategies. Some examples include complex topographies~\cite{volpe2017topography}, non-uniform distributions of targets~\cite{wosniack2017evolutionary}, the presence of gradient forces or obstacles, or even self-interacting foragers~\cite{barbier2022self}.  Another interesting perspective involves endowing the agents with a memory which goes beyond a mere step counter. For instance, the agent could remember its past decisions for a given number of steps, similar to what is considered in~\cite{meyer2021optimal}. Moreover, RL can be used to investigate multi-agent scenarios~\cite{Lopezincera20,Lopezincera21} and their connection with known collective phenomena such as flocking~\cite{santos2009can, gosztolai2019collective}.

\begin{acknowledgments}
We acknowledge Michele Caraglio and Harpreet Kaur for insightful discussions. This work was supported by the Austrian Science Fund (FWF) through the SFB BeyondC F7102, and the European Research Council (ERC, QuantAI, Project No. 10105529). A.L and H.J.B also acknowledge support by the Volkswagen Foundation (Az:97721). G.M-G also acknowledges funding from the European Union.
Views and opinions expressed are however those of the author(s) only and do not necessarily reflect those of the European Union, the European Research Council or the European Research Executive Agency . Neither the European Union nor the granting authorities can be held responsible for them. Animal icons used in \cref{fig:fig0} were adapted from \href{https://www.svgrepo.com}{www.svgrepo.com}.
\end{acknowledgments}

\bibliographystyle{unsrt}
\bibliography{biblio}

\onecolumngrid 

\newpage

\appendix
\label{appendix}

%__________________BENCHMARK MODELS_____________________
 
\section{Benchmark models} \label{app:benchmark_models}
This appendix defines the model families used as benchmarks. These include i) the family of L\'evy distributions~\cite{clauset2009power},
\begin{equation}
\label{eq:LW_dist}
    \Pr(L)=\zeta^{-1}_{(1+\beta, 1)} L^{-1-\beta}\, , 
\end{equation}
where $\zeta_{(1+\beta, 1)}=\sum_{\ell=0}^\infty (\ell+1)^{-1-\beta}$ is the Riemann zeta function, and  ii) the family of discrete bi-exponential distributions~\cite{clauset2009power} 
\begin{equation}
\label{eq:biexp_dist}
\Pr(L) = \sum_{i=1,2} \omega_i (1-e^{-1/d_i}) e^{-(L-1)/d_i} \, ,
\end{equation}
where $d_i$ are length scales and the mode weights satisfy $\sum_{i=1,2} \omega_i=1$. Note that in each case we consider that the minimum step length is $L=1$.

The search efficiency (\cref{eq:efficiency}) of benchmark models and trained policies are computed the same way. To this end, the step-length distribution of a benchmark model is translated into an equivalent, fixed policy using \cref{eq:solution}, and then the average search efficiency is obtained from $10^4$ walks, each consisting of $T=20000$ small steps of length $d=1$.

\begin{figure}[htb!]
\centering
\includegraphics[width=\textwidth]{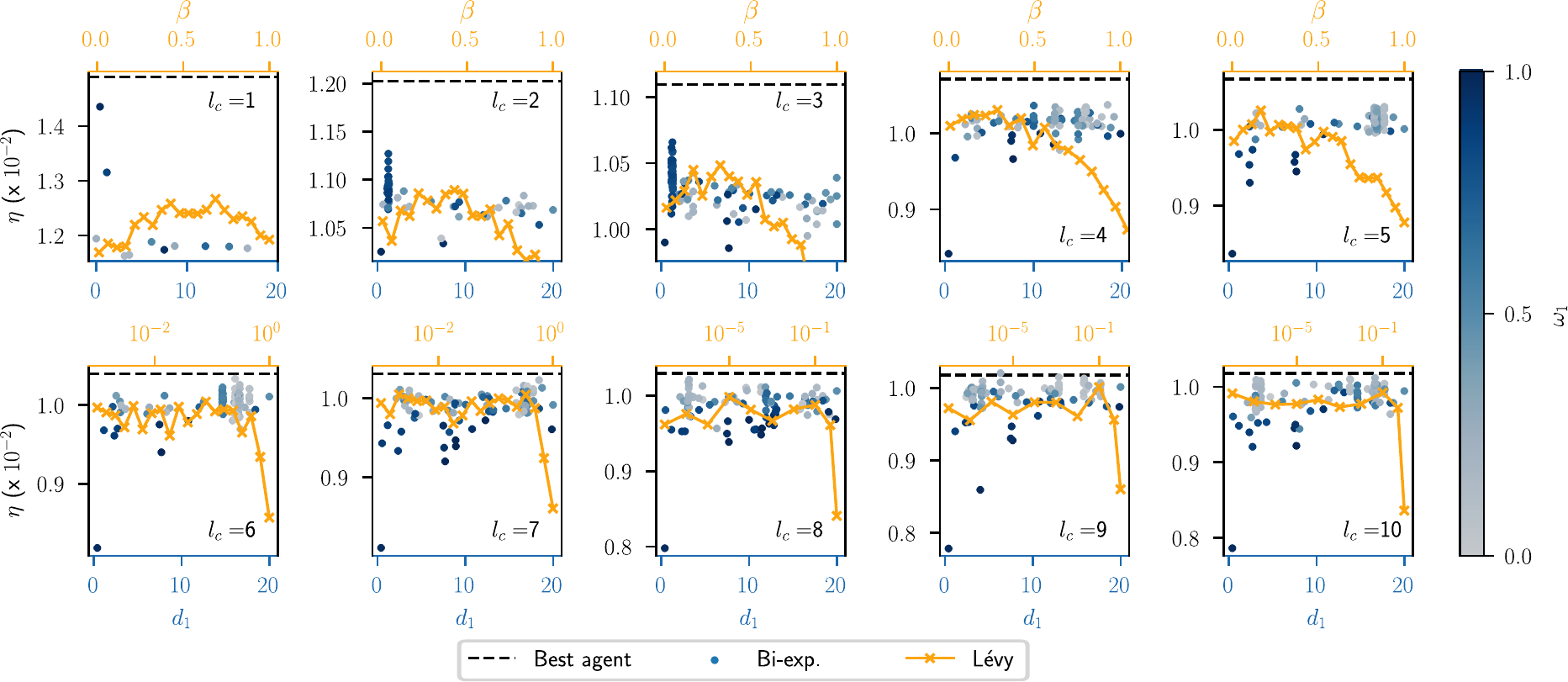}
\caption{Average search efficiency ($\eta$) achieved by agents with step-length distributions given by \cref{eq:LW_dist} (L\'evy) and \cref{eq:biexp_dist} (Bi-exponential), for different values of their model parameters, in environments with different cutoff length $l_\mathrm{c}$. We use this search to find the L\'evy and bi-exponential distributions that achieve the highest efficiency to benchmark the RL agent's performance. In all panels, each point shows the average over $10^4$ walks. In panels with $l_\mathrm{c}=1,3$, $d_2=10^5$. In panels with $l_\mathrm{c}=2,4,5,6,...,10$, $d_2=10^2$. Dashed lines indicate the efficiency of the best RL agent in each case.}
\label{fig:benchmark_results}
\end{figure}

With a method to calculate the average search efficiency at hand, the model parameters are optimized in order to maximize the average search efficiency for both model families and for each environment (differing in cutoff length). We use Bayesian optimization for the family of bi-exponential distributions and grid searches for the L\'evy distributions, via the Tune Python package~\cite{liaw2018tune}, to obtain the model parameters that optimize the performance of each of the benchmark models for the environment under consideration. \cref{fig:benchmark_results} shows the performance of each of the benchmark models for different values of the model parameters. We take the distributions that achieve the highest search efficiency for benchmarking. Examples of such optimizations are presented in the Tutorials section of the accompanying repository~\cite{ourRepo}.

\begin{figure}[t]
    \includegraphics[width=0.5\textwidth]{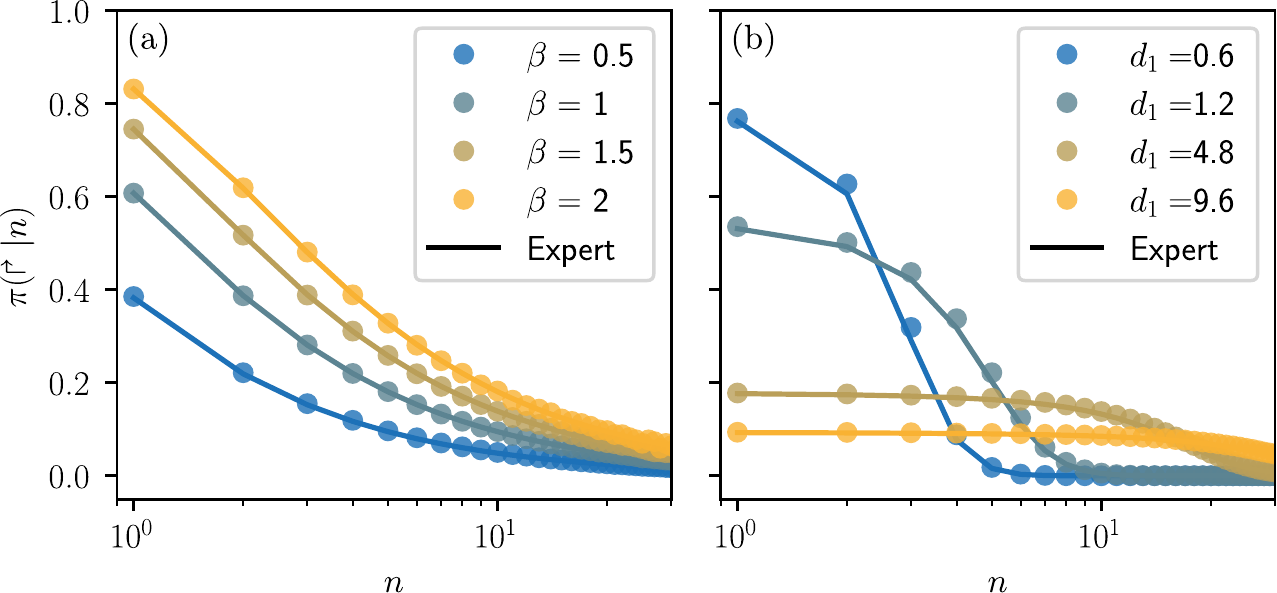}
    \caption{Comparison of benchmark policies (solid lines) with the policies (colored dots) of a PS agent using imitation learning. Panel a) shows L\'evy distributions with varying exponent $\beta$ and panel b) shows bi-exponential distributions with varying $d_1$, $w_1=0.94$ and $d_2 =5000$ (see~\cref{app:benchmark_models} for details).}
    \label{fig:fake_training}
\end{figure}

%__________________PS_____________________
\section{Projective simulation}\label{app:PS}
We use projective simulation (PS)~\cite{briegel2012projective, Mautner15, Makmal16, Melnikov17} to model the RL agent and its learning process. PS agents process the environmental information, that is, decide and learn by means of an episodic memory (ECM), which is mathematically described by a graph.

In the simplest case, the nodes of this graph are distributed in two layers, one of which represents the states and the other the actions. Each state is connected to all possible actions by directed, weighted edges. The decision making  of a PS agent is stochastic: the probability of performing action $a \in \mathcal{A}$ having perceived a state $s \in \mathcal{S}$ is
\begin{equation}
	\label{eq:ps_policy}
	\pi(a|s) = \frac{h(s,a)}{\sum_{a'\in \mathcal{A}} h(s,a')} \, ,
\end{equation}
where $h(s,a)$, called \hvalue, is the weight of the edge that connects $s$ with $a$. The agent learns by updating the \hvalues--and thus, its policy $\pi(a|s)$--at the end of each RL step. During one interaction with the environment, i.e., one RL step, the agent (i) perceives a state $s$, (ii) decides on an action $a$ by sampling from its current policy $\pi(a|s)$ and performs the action, (iii) receives a reward $R$ from the environment, and (iv) updates its memory accordingly. The latter is implemented by updating the matrix $\mathcal{H}$ that contains the \hvalues:
\begin{equation}
\label{eq:h_update}
    \mathcal{H} \leftarrow \mathcal{H} - \gamma (\mathcal{H}-\mathcal{H}_0) + \mathcal{G} R\,,
\end{equation}
where $\gamma$ is a damping factor that controls how quickly the agent forgets, $\mathcal{H}_0$ is the initial $\mathcal{H}$ matrix and \gmatrix~is the so-called glow matrix, which allows the agent to learn from delayed rewards, i.e. rewards that are only received several interactions with the environment later. The glow mechanism tracks which edges in the ECM were traversed prior to getting a reward. To do so, whenever there is a transition from state $s$ to action $a$, its corresponding edge ``glows'' with a certain intensity given by the glow value $g(s,a)$, which is stored in the \gmatrix~matrix. The glow has an initial value of $g(s,a)=0$ and increases by 1 every time the edge is traversed (for details on the definition and update of the \gmatrix~matrix, see \cite{Mautner15, Boyajian19}). At the end of each RL step, \gmatrix~is damped as
\begin{equation}
   \mathcal{G} \leftarrow (1-\eta_g) \mathcal{G}, 
   \label{app:glow}
\end{equation}
where $\eta_g$ is the damping factor. Both $\eta_g$ and $\gamma$ are considered hyperparameters of the model and are adjusted to obtain the best learning performance.

%__________________IMITATION_____________________
 
\subsection{Imitation learning of benchmarks} \label{app:imitation}
In this section, we study whether a PS agent is actually able to learn known foraging strategies within the proposed framework. To do so, we translate the original foraging problem into an imitation problem. Imitation learning is commonly used to simplify reinforcement learning problems with sparse rewards. Instead of learning in the original, complex scenario, the RL agent is trained to imitate an \emph{expert} already equipped with the optimal policy. The expert is then considered to perform only succeeding trajectories, i.e. sets of actions that lead to rewarded states. In this way, the reward sparsity of the original problem can be avoided and learning the optimal policy is greatly simplified. Nonetheless, the RL agent is still required to correctly update its policy, hence imitation learning is a useful experiment to understand if the proposed framework is adequate for an agent that needs to learn optimal foraging strategies.

In what follows, we consider environments for which we assume that certain distributions $\Pr(L)$ are optimal. This means that the expert's policy is calculated inserting $\Pr(L)$ in \cref{eq:solution}. In the imitation scheme, one RL step proceeds as follows:
\begin{enumerate}
    \item a step of length $L$ is sampled from $\Pr(L)$;
    \item the counter $n$ of the RL agent is set to $N$ such that $Nd=L$;
    \item the agent RL performs the turning action, hence having effectively done a step of exactly length $L$;
    \item a reward $R=1$ is given to the agent after the action ($\Rsh$) in state ($N$);
    \item the agent updates its policy via \cref{eq:h_update} based on this reward;
\end{enumerate}

In~\cref{fig:fake_training}a-b we show the convergence of the policy of an RL agent trained to imitate an expert equipped with policies calculated from L\'evy and bi-exponential distributions with multiple parameters. As shown, the agent's policy correctly converges to the expert's. This ensures that the proposed framework, together with the PS update rule, can adequately accommodate the most typical foraging strategies, if these were to be optimal in the given environment.

%__________________LEARNING_____________________
 
\subsection{Details about the training and the model parameters of the agent} \label{app:learning}
The RL learning process of the PS agent consists of $N_{\mathrm{ep}}\simeq 10^4$ episodes, each of which contains $T$ RL steps. The agent's memory has a two-layer structure of states and actions, as described in the previous section. Its learning mechanism is governed by \cref{eq:h_update}. There are $T$ states, one for each possible value that the counter $n$ can in principle take before the episode is over; and 2 actions, continue in the same direction or turn. At the beginning of each episode, the \gmatrix~matrix of the agent is reset and both the agent's position and the positions of the targets are randomly sampled. The agent keeps updating its policy until $N_{\mathrm{ep}}$ episodes are completed. In order to accurately assess search efficiency at different stages of training (\cref{fig:efficiency}), we perform a post-training evaluation: we take the policy in a given episode, freeze it, and have the agent employ this policy to perform $10^4$ walks consisting of $20000$ small steps of length $d=1$ in environments with different target distributions. We train 10 independent agents in each scenario (given by different cutoff lengths) and consider that the best agent is the one that achieves the highest average search efficiency in the post-training evaluation using the policy after the last episode. 

The learning parameters $\gamma$ and $\eta_g$ were adjusted in order to achieve the best performance in each case. The values of all parameters for each scenario are given in \cref{table:learning_par}. Examples of trainings are presented in the accompanying repository~\cite{ourRepo}.

\begin{table}[htb!]
\centering
\begin{tabular}{c|c|c|c|}
\cline{2-4}
 & \cref{fig:fake_training,fig:evolution_learning}c         & \cref{fig:efficiency,fig:chosen_policy,fig:policies_and_dist}         & \cref{fig:evolution_learning}a-b         \\ \hline
\multicolumn{1}{|c|}{$\pi_0 (\uparrow|n)$} & 0.5       & 0.99      & 0.5       \\ \hline
\multicolumn{1}{|c|}{$N_{\mathrm{ep}}$}                  & 1000      & 12000     & 69000     \\ \hline
\multicolumn{1}{|c|}{$T$}                       & 10000     & 20000     & 3000      \\ \hline
\multicolumn{1}{|c|}{$\eta_g$}                    & $10^{-7}$ & 0.1       & 0.1       \\ \hline
\multicolumn{1}{|c|}{$\gamma$}                  & 0         & $10^{-5}$ & $10^{-6}$ \\ \hline
\end{tabular}
\caption{Training parameters used to obtain the data presented in the respective figures.
}
\label{table:learning_par}
\end{table}

\subsection{Policy initialization and finite training times} \label{app:policy_init}
\begin{figure}[h!]
\centering
\includegraphics[width=0.6\textwidth]{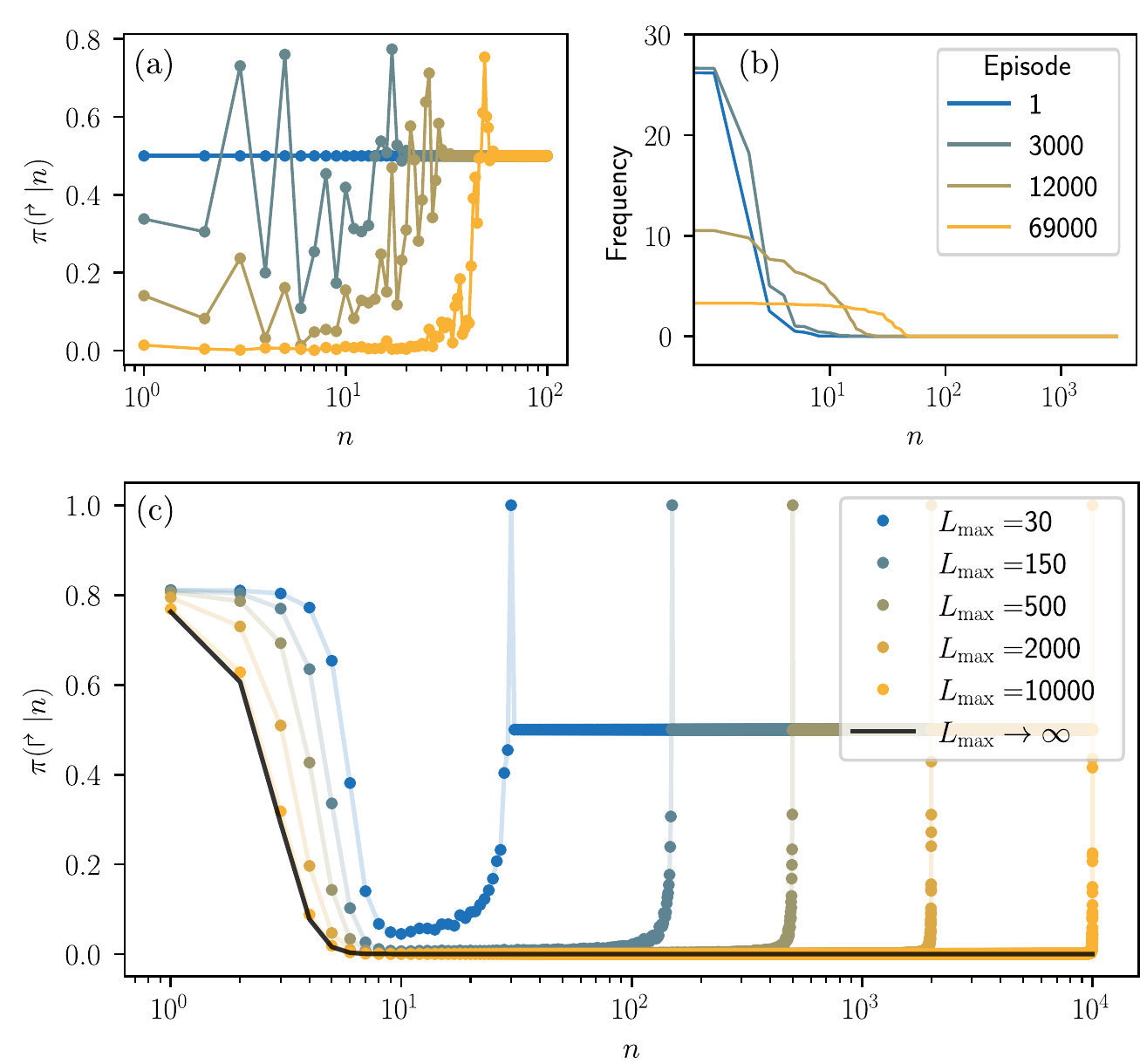}
\caption{a) Evolution of the policy throughout the learning process for one agent initialized with $\pi_0 (\Rsh|n) = 0.5$ for all $n$. b) Frequency (in percentage of RL steps) with which the agent perceives each of the states $n$ at a given episode (the color code for each episode is the same in panels a and b). Parameters are  $l_\textrm{c}=10$, $d=1$, $r=0.5$, and $\rho=0.01$. c) Learned policies for imitation learning (for details, see \cref{app:imitation}). The policy that corresponds to the distribution without cutoff ($L_{\mathrm{max}}\rightarrow \infty$) is shown for reference. Parameters are $d_1=0.6$, $d_2=5000$, and $\omega_1=0.94$.}
\label{fig:evolution_learning}
\end{figure}

In this section, we motivate our choice of policy initialization.
In the literature, the policy of a PS agent is typically initialized such that $\pi_0 (\uparrow|n) = \pi_0 (\Rsh|n) = 0.5$ holds for all $n$. While $\pi_0(\Rsh|n) =0.01 $ is used in the main text, in this appendix, we first consider the $\pi_0 (\Rsh|n) = 0.5$ policy initialization instead.  That is, the agent initially chooses randomly and uniformly between the two possible actions $\uparrow$ and $\Rsh$. In PS, this is enforced by setting all $h$-values in the initial $\mathcal{H}_0$ matrix to $h(s,a)=1$ $\forall s, \, a$. As we see in \cref{fig:evolution_learning}a, the agent progressively learns to decrease $\pi(\Rsh|n)$, which in turn leads to longer steps, and to the agent reaching larger values of $n$ as the learning  progresses. Nonetheless, reaching large $n$ is still exponentially costly. Moreover, to properly train the policy at a given $n$, the agent needs to first reach such value and then sample enough rewards. In \cref{fig:evolution_learning}b we show the frequency with which $n$ is observed in different episodes. As expected, due to the update of the policy, the probability of reaching larger $n$ increases as the training progresses. However, this training requires large computation times and still the agent does not sample steps longer than $L=40$.

As can be seen in \cref{fig:evolution_learning}a, even after $69000$ episodes (yellow line), the policy exhibits a decay back to initialization values for large $n$. The policy $\pi(\Rsh|n)$ at a given $n$ depends not only on the received rewards at such $n$, but also on the rewards that will be attained for $n'>n$ due to the glow (\cref{app:glow}). Thus, not reaching $n'>n$ also hampers the policy update of $\pi(\Rsh|n)$. 

Although imitation learning bypasses this problem (see \cref{app:imitation}), it can still be used to further understand the effect of the problem on the trained policies. In \cref{fig:evolution_learning}c, we consider an agent that learns to imitate a bi-exponential distribution that has a cutoff at $L_{\mathrm{max}}$, such that $\Pr(L\geq L_{\mathrm{max}}) = 0$. As we see, the policy decays at $n\sim L_{\mathrm{max}}$. Moreover, the smaller $L_{\mathrm{max}}$, the more the policy at small $n$ deviates from that of the bi-exponential distribution without cutoff ($L_{\mathrm{max}}\rightarrow \infty$).

To improve the exploration of longer steps, instead of increasing the training time, we change the initialization of the policy so that the agent takes longer steps more frequently already at the beginning of the training. In particular, throughout the numerical experiments presented in the main text, we consider that $\pi_0 (\Rsh|n) = 0.01$ for all $n$, which is implemented by setting the $h$-values in the initial $\mathcal{H}_0$ matrix to $h(s,\uparrow)=0.99$, and $h(s,\Rsh)=0.01$ $\forall s$.

\section{Efficiency advantage of turning at $nd\simeq l_\mathrm{c}$}
\label{app:geometry}
In this appendix, we outline a geometric argument which supports the observation that learned policies show an increase in the probability of turning after $nd \simeq l_\mathrm{c}$ steps.

\begin{figure}[h]
   \includegraphics[width=0.2\columnwidth]{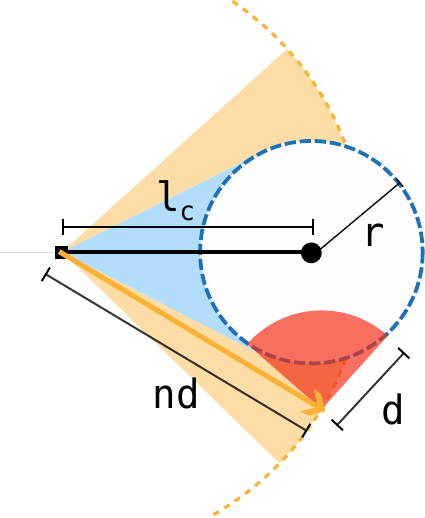}
   \caption{Schematic illustration of the benefit of turning after $nd\simeq l_\mathrm{c}$ steps in a simplified scenario with only a single target, the center of which is represented with a black dot and which has a detection radius $r$. After the detection of the target, the agent is displaced by a distance $l_\mathrm{c}$ from the target (black line). Then, we assume that it continues to walk straight for $nd\simeq l_\mathrm{c}$ at a random angle (orange line). In this case, the blue area marks the angle range within which the agent returns to the target when walking straight. The yellow area indicates the angle range at which the target is missed but is still within reach (after one step) if the agent turns by a suitable angle (red area).}
    \label{fig:angles}
\end{figure}

Let $l_\mathrm{c}$ be much smaller than $\sqrt{\rho^{-1}}$ (i.e. much smaller than the mean spacing between targets), and consider the following situation: an agent has detected a target and has just been displaced by $l_\mathrm{c}$. Then, within a radius of $l_\mathrm{c}$ of the agent, there is typically no other target than the one previously detected. Let us therefore assume that the only target nearby is the one previously detected. Then, the environment for the next $nd\simeq l_\mathrm{c}$ steps of the agent only contains the previously detected target, see Fig.~\ref{fig:angles}. We write $nd\simeq l_\mathrm{c}$ rather than $nd =  l_\mathrm{c}$ because $l_\mathrm{c}$ does not have to be an integer multiple of $d$. $n$ is chosen such that it is possible to return to the target and to detect it by walking straight for precisely $n$ steps, $n = \lceil \frac{l_\mathrm{c}-r}{d}\rceil$.

We consider two probabilities: (i) the probability $p_{\uparrow}$ that the agent returns directly to the target by walking straight for $nd\simeq l_\mathrm{c}$, and (ii) the probability $p_{\Rsh}$ that the agent returns to the target by first walking straight for $nd$ (without detecting the target) and then walking for one step of length $d$ at a random angle (it turns). It is a matter of simple trigonometric considerations to compute these probabilities. $p_{\Rsh}$ is plotted in the inset of \cref{fig:policies_and_dist}a. We find numerically that $p_{\Rsh}$ is even larger than $p_{\uparrow}$ for the values of $l_\mathrm{c}$ plotted in the inset of \cref{fig:policies_and_dist}a.

\end{document}